\documentclass[10pt,preprint]{aastex}
\begin{document}

\title{THE [O III] VEIL: ASTROPAUSE OF ETA CARINAE'S WIND?\altaffilmark{1}}
%% or
%% \title{THE [O III] VEIL AROUND ETA CARINAE: A TERMINAL SHOCK?\altaffilmark{1}}

\author{Nathan Smith\altaffilmark{2}}
\affil{Center for Astrophysics and Space Astronomy, University of
Colorado, 389 UCB, Boulder, CO 80309}

\author{Jon A.\ Morse}
\affil{Department of Physics and Astronomy, Arizona State University,
Box 871504, Tempe, AZ 85287-1504}

\author{John Bally\altaffilmark{3}}
\affil{Center for Astrophysics and Space Astronomy, University of
Colorado, 389 UCB, Boulder, CO 80309}

\altaffiltext{1}{Based in part on observations made with the NASA/ESA
{\it Hubble Space Telescope}, obtained at the Space Telescope Science
Institute, which is operated by the Association of Universities for
Research in Astronomy, Inc., under NASA contract NAS5-26555.}

\altaffiltext{2}{Hubble Fellow; nathans@casa.colorado.edu}

\altaffiltext{3}{Visiting Astronomer, Cerro Tololo Inter-American
Observatory, National Optical Astronomy Observatory, operated by the
Association of Universities for Research in Astronomy, Inc., under
cooperative agreement with the National Science Foundation.}

\begin{abstract}

We present narrowband images of $\eta$~Carinae in the light of [O~{\sc
iii}] $\lambda$5007 obtained with the {\it Hubble Space
Telescope}/Wide Field Planetary Camera 2 ({\it HST}/WFPC2), as well as
a ground-based image in the same emission line with a larger field of
view.  These images show a thin veil of [O~{\sc iii}] emission around
$\eta$~Car and its ejecta, confirming the existence of an
oxygen-bearing ``cocoon'' inferred from spectra.  This [O~{\sc iii}]
veil may be the remnant of the pre-outburst wind of $\eta$~Car, and
its outer edge probably marks the interface where $\eta$ Car's ejecta
meet the stellar wind of the nearby O4~V((f)) star HD~303308 or other
ambient material in the Carina nebula -- i.e., it marks the
``astropause'' in $\eta$ Car's wind.  This veil is part of a more
extensive [O~{\sc iii}] shell that appears to be shaped and ionized by
HD~303308.  A pair of {\it HST} images with a 10 yr baseline shows no
significant proper motion, limiting the expansion speed away from
$\eta$~Car to 12$\pm$13 km s$^{-1}$, or an expansion age of a few
times 10$^4$ yr.  Thus, this is probably the decelerated pre-outburst
LBV wind of $\eta$ Car.  The [O~{\sc iii}] morphology is very
different from that seen in [N~{\sc ii}], which traces younger dense
knots of CNO-processed material; this represents a dramatic shift in
the chemical makeup of material recently ejected by $\eta$~Car.  This
change in the chemical abundances of the ejecta may have resulted from
the sudden removal of the star's outer envelope during $\eta$ Car's
19th century outburst or an earlier but similar event.

\end{abstract}


\section{INTRODUCTION}

Eta Carinae is our best example of an extremely massive star
surrounded by material recently ejected during the late stages of
evolution.  Its astonishingly complex circumstellar ejecta exhibit a
diverse range of properties with nebulae nested inside one another,
having resulted from repeated outbursts of the star: the ``Weigelt
knots'' within a few 10$^2$ AU of the star and a dynamical age of
60-110 yr (Dorland et al.\ 2004; Smith et al.\ 2004; Weigelt et al.\
1995; Hofmann \& Weigelt 1988); the bipolar ``Little Homunculus'' with
a size of several 10$^3$ AU and an age of $\sim$110 yr (Ishibashi et
al.\ 2003; Smith 2005); the neutral and dusty bipolar ``Homunculus''
with an extent of 0.2 pc and total mass exceeding 10 M$_{\odot}$
(Smith et al.\ 2003) ejected during the Great Eruption of 1843 (Morse
et al.\ 2001; Currie \& Dowling 1999; Smith \& Gehrz 1998); and
finally the nitrogen-rich, ionized ``outer ejecta'' found outside the
Homunculus, some of which were ejected during the Great Eruption, and
some of which may be decades to centuries older (Walborn et al.\ 1978;
Walborn \& Blanco 1988; Morse et al.\ 2001; Weis 2001).  These older
ejecta trace the mass-loss history of $\eta$~Car, and may offer clues
to its evolutionary state and the nature of its stellar wind before
the Great Eruption.

At what separation from the star can material no longer be attributed
directly to ejecta from $\eta$ Car or the influence of its stellar
wind?  Bohigas et al.\ (2000) have proposed the existence of an old
bipolar shell ejected by $\eta$~Car, and Smith (2002a) presented
circumstantial evidence that the shape of the Keyhole nebula may have
been influenced by $\eta$~Car's wind.  If enhanced mass loss occurred
during a luminous blue variable (LBV) phase that lasted $\sim$10$^5$
yr with $v_{\infty}\simeq$500 km s$^{-1}$, for example, then the
products of $\eta$ Car's outbursts may potentially be seen 1$\arcdeg$
away or more ($\sim$50 pc).  Before reaching such distances, however,
ejecta from $\eta$~Car interact with stellar winds and radiation from
several of the other luminous early-type stars in the Carina nebula
(e.g., Walborn 1973, 1995; Feinstein 1995).  This interaction might
make the material unrecognizable as stellar ejecta from $\eta$~Car on
dynamical grounds, leaving chemical abundances of the diluted material
as the only potential diagnostic.

Smith \& Morse (2004) presented spectra of the outer ejecta of $\eta$
Car, showing that while ejecta immediately outside the Homunculus were
already known to be nitrogen rich and severely depleted of oxygen
(Davidson et al.\ 1982; Dufour et al.\ 1997), [O~{\sc ii}] and [O~{\sc
iii}] lines became stronger with increasing separation from the star.
In that paper we proposed that the nitrogen rich condensations ejected
at high speed by $\eta$~Car are overtaking a normal-composition
``cocoon'' deposited by previous stellar-wind mass loss, and that this
interaction gives rise to the soft X-ray shell around $\eta$~Car
(Corcoran et al.\ 1995, 1999, 2004; Seward et al.\ 2001; Weis et al.\
2004).  In this paper we present [O~{\sc iii}] images of $\eta$~Car's
environment, which we interpret as confirmation of this scenario.

\section{OBSERVATIONS}

We obtained images of $\eta$ Car, the Keyhole, and portions of the
surrounding Carina nebula with 0$\farcs$8 seeing on 2003 March 10
using the Cerro Tololo Inter-American Observatory (CTIO) 4m telescope.
The 8192$\times$8192 pixel imager MOSAIC2, which has a 2$\times$4
array of 2048$\times$4096 pixel CCDs, provided a 35$\farcm$4 field of
view with a pixel scale of 0$\farcs$26.  The images reported here were
obtained with a narrowband ($\sim$80 \AA-wide) filter centered on
[O~{\sc iii}] $\lambda$5007.  Several individual 60 sec exposures were
obtained with slight positional offsets to fill the inter-chip gaps
and to correct for severe CCD blooming by $\eta$~Car, with a total
exposure time of 300 sec over most of the observed area.  Images were
reduced in the standard fashion with the MSCRED package in IRAF, and
absolute sky coordinates were computed with reference to USNO catalog
stars.  A section of the CTIO image showing the environment around
$\eta$~Car is displayed in Figure 1.

We also observed $\eta$~Car on 2003 August 8 (program GO-9775) with
the Wide Field Planetary Camera 2 (WFPC2) aboard the {\it Hubble Space
Telescope} ({\it HST}) through the F502N filter to capture [O~{\sc
iii}] $\lambda$5007 emission from the circumstellar environment.  We
compared these data with similar F502N images from the {\it HST}
archive obtained on 1993 December 31 to search for temporal flux
variations and proper motions of nebular structures. Table~1 contains
a log of the WFPC2 F502N observations. Each data set consisted of
several individual exposures with a range of integration times.  All
observations had $\eta$~Car centered on the WF3 chip. We used the
shorter exposures to patch pixels that were saturated or affected by
bleeding from the bright central star of $\eta$~Car in longer
exposures.  We combined multiple exposures to reject cosmic rays, as
we have done with our previous {\it HST} imaging of $\eta$~Car (Morse
et al.\ 1998, 2001; Smith et al.\ 2000, 2004), and we corrected the
images for the geometric distortion of the WFPC2 instrument using
standard routines in IRAF/STSDAS.  Since we are interested in
examining proper motions of nebular features over the $\sim$10 yr
interval between the two observations, we needed to carefully register
the two epochs.  Simple alignment with rotation and mosaicing using
the WMOSAIC task in STSDAS does not give sufficiently accurate
registration, so we used $\sim$15 field stars as tie points to
establish a common coordinate frame (see also Morse et al.\ 1998,
2001; Fesen et al. 2001).  Using a second-order polynomial fit, the
images were aligned to $\sim$0.1 pixel rms.  The final F502N images
from 1993 and 2003 are shown in Figure 2.

\section{THE [O~{\sc iii}] VEIL}

Figure 1 shows the environment around $\eta$~Car in [O~{\sc iii}]
$\lambda$5007, revealing different morphology from that seen in other
emission line tracers like [N~{\sc ii}] (e.g., Thackeray 1949, 1950;
Gaviola 1950; Walborn 1976; Meaburn et al.\ 1987, 1993, 1996; Morse et
al.\ 1998; Weis 2001).  Instead of the dense knots and filaments of
the nitrogen-rich outer ejecta, Figure 1 shows a thin limb-brightened
veil of [O~{\sc iii}] emission encasing the Homunculus with a radius
of about 30--40\arcsec.\footnote{While the S Condensation, the S
Ridge, and the NN ``jet'' are seen clearly in Figure 1, this is due to
starlight scattered by dust, since these features show no detectable
[O~{\sc iii}] emission in spectra (Smith \& Morse 2004).}  This veil
has a thin, limb-brightened outer edge best seen northwest of the
Homunculus, especially in {\it HST} images of the same emission line
(Fig.\ 2).  The difference between [O~{\sc iii}] and [N~{\sc ii}]
emission is striking in the color representation of Figure 3.  No
clear outer edge to the [O~{\sc iii}] veil is seen south of the
Homunculus, which is probably related to the lack of soft X-ray
emission in that direction (Seward et al.\ 2001).  The southern part
of the veil may also be missing because of a lack of ionizing photons
(see below).

Smith \& Morse (2004) proposed the existence of a normal-composition
``cocoon'' surrounding $\eta$~Car; as the nitrogen-rich outer ejecta
run into this cocoon, their chemical abundances are modified to
include varying amounts of oxygen.  Independent from its excitation
mechanism, the [O~{\sc iii}] emission in Figures 1 and 2 indicates a
smooth distribution of gas with a significant amount of oxygen and
with a different morphology from that of the nitrogen-rich ejecta (see
Fig.\ 3$b$).  This indicates that there is indeed oxygen surrounding
the outer ejecta, confirming the existence of the ``cocoon''.

Figure 1 also shows a larger shell structure, seen as a thin filament
adjoining the southeastern corner of the ``veil'', extending toward
the east for about 1$\farcm$5, and then turning northward and looping
back toward the Keyhole nebula.  Part of this shell structure was seen
by Bohigas et al.\ (2000), but they interpreted this as part of a
large bipolar structure with a polar axis running northeast to
southwest -- perpendicular to that of the Homunculus.  This bipolar
shape is not confirmed by our new images.  In our higher resolution
image, this structure gives the impression of an expanding bubble
whose progress was thwarted toward the south by the ejecta around
$\eta$~Car; it looks as though a veil were draped over the Homunculus
as $\eta$ Car and its dense ejecta punch into the shell.  This shell
and veil may mark the interaction of stellar winds and dense ejecta.
Interestingly, the very hot and massive O4~V((f)) star HD~303308
(Walborn 1973; Walborn et al.\ 2002), which should have a powerful
stellar wind and strong UV radiation, is seen projected inside this
bubble.

HD~303308 may be partly responsible for the appearance of this shell,
as noted by Bohigas et al.\ (2000).  Despite being one of the most
luminous stars in the Galaxy, $\eta$~Car itself is a pitiful source of
ionizing radiation because its dense stellar wind and the dusty
Homunculus nebula trap the UV output of the central star.  However,
O$^{++}$ requires a significant flux of photons above 54 eV.  Thus,
the [O~{\sc iii}] emission in Figure 1 requires some other source for
its ionization and excitation.  [O~{\sc iii}] emission is brightest on
the northern side of the cocoon around $\eta$~Car, pointing toward
radiation or the stellar wind from HD~303308 as the most likely
culprit.  Shocks could excite the [O~{\sc iii}] line, but this is
unlikely due to the slow velocity of the veil, unless it is a standing
shock (see below).  Furthermore, optical spectra of the ``W Arc'' (the
thin filament at the western edge of the veil around $\eta$~Car; Smith
\& Morse 2004) and part of the distant [O~{\sc iii}] shell (Bohigas et
al.\ 2000) show line ratios typical of photoionized gas in an H~{\sc
ii} region, but with higher densities than surrounding regions of the
Carina nebula.

Even if the [O~{\sc iii}] emission is not currently dominated by shock
excitation, the veil in Figures 1 and 2 may trace a photoionized
density enhancement or contact discontinuity in a standing shock
front, resulting from the interaction of dense ejecta and a stellar
wind.  For example, suppose that leading up to the Great Eruption in
the 1840's (or more likely, leading up to the last major eruption
before the 19th century outburst) $\eta$~Car had a dense stellar wind,
comparable to its extreme present-day wind but with roughly solar
composition.  This would be the case for the stellar wind before CNO
ashes first made their way to the surface of the star.  At some point,
this dense wind would interact with the faster and more rarefied
stellar winds from massive O-type stars in the Trumpler 16 cluster,
like HD~303308.  If this scenario applies, then the [O~{\sc iii}] veil
seen in Figures 1 and 2 is the terminal shock, or ``astropause'' of
$\eta$~Car's stellar wind.\footnote{We use the term ``astropause'',
not to imply any similarity to the physical conditions of the
heliopause in our solar system, but simply to denote that this is the
farthest outer boundary of $\eta$ Car's undiluted stellar wind.}

\section{HST IMAGES AND PROPER MOTIONS}

In addition to showing the structure of the [O~{\sc iii}] veil in more
detail than ground-based images, our multi-epoch {\it HST} images
taken with the same filter and instrument can also constrain the
motion of the shell.  With a time baseline between the 1993 and 2003
WFPC2 images of 9.6 yr, we can measure the positions of nebular
features to an accuracy of 10-20\% of a WF pixel (0$\farcs$1),
sensitive to motion comparable to the sound speed in ionized gas.

An efficient way to search for movement is to subtract registered
frames at two epochs (e.g., Morse et al.\ 2001), so that moving
condensations show adjacent light and dark features.  This is shown in
Figure 3$a$ for the WF3 images in the F502N filter.  The expansion of
the Homunculus is clear in Figure 3$a$, as is the movement of the NN
jet and S condensation.  In all three of these cases, the moving
material is fast ($v>100$ \ km s$^{-1}$) and is seen as dust-scattered
continuum light in the F502N filter, instead of [O~{\sc iii}]
emission.  Figure 3$a$ shows no sign of proper motion in the veil,
where all the thin filamentary [O~{\sc iii}] emission has subtracted
to better than the residual noise.  We find essentially no detectable
motion nor surface brightness variation of the [O~{\sc iii}] veil
during the $\sim$10 yr time interval.  The lack of any discernable
proper motion is obvious when one blinks between the images at the two
epochs on a computer monitor.  If the outer veil is expanding at all,
it is moving very slowly compared to the ejecta closer to $\eta$ Car.

Figure 4 shows an example of intensity tracings through the W Arc in
the 1993 and 2003 F502N images.  The limb-brightened edge of the W Arc
corresponds to the strong feature at $\sim$33\arcsec\ from the star
(see Fig.\ 2$b$).  Again, the [O~{\sc iii}] veil shows no proper
motion.  To quantify this, we measured the positions of the W Arc at
both epochs from the tracings in Figure 4 using five different
methods: a flux-weighted centroid, cross-correlation, and Gaussian,
Lorentzian, and Voigt profile fits.  We measured proper motion of
+11$\pm$12 mas, or a tangential velocity of 12$\pm$13 km s$^{-1}$ away
from the star in the plane of the sky for a distance of 2250 kpc
(Smith 2002$b$).  This measurement uncertainty is comparable to the
uncertainty in the registration of the two images, confirming that we
do not detect any significant motion of the [O~{\sc iii}] veil.  The
dynamical age of the [O~{\sc iii}] veil is at least 3$\times$10$^4$
yr, but the true age of the veil could obviously be less if it has
decelerated from its initial speed.  Thus, our data give little
information on the actual age and origin of the [O~{\sc iii}] veil,
except that it is consistent with being the dense swept-up wind lost
during a pre-outburst LBV phase.

The slow motion of the [O~{\sc iii}] veil also gives important clues
about its excitation mechanism.  Its speed must be close to the sound
speed in ionized gas.  Shock velocities $\ga$100 km s$^{-1}$ are
normally required to account for [O~{\sc iii}] emission (e.g.,
Hartigan et al.\ 1999), implying that the veil is dominated by
photoionization.  The exception to this would be a stationary standing
shock in colliding winds, but this would only make sense in the region
directly between $\eta$ Car and HD~303308, not on the east and west
sides of the veil.  Additionally, no X-ray emission has been reported
from the region between $\eta$ Car and HD~303308, except for the soft
X-ray shell around $\eta$ Car, which is {\it inside} the veil. Thus,
we conclude that the most likely excitation mechanism for the [O~{\sc
iii}] veil is external UV irradiation by HD~303308.

\section{ETA CAR'S MASS-LOSS HISTORY}

Based on spectra of several positions in $\eta$~Car's outer ejecta, we
proposed a shift in the nebular abundance pattern -- from
nitrogen-rich CNO ashes near the star to more normal composition
material farther away -- and we suggested that this is the result of
nitrogen-rich ejecta running into slower normal-composition material
(Smith \& Morse 2004).  The [O~{\sc iii}] images we have presented
here show a veil of bright oxygen emission, with a different and
distinct morphology from that seen in nitrogen images (Fig.\ 3$b$);
the [N~{\sc ii}] condensations are found inside the boundaries of the
[O~{\sc iii}] veil.  We interpret this [O~{\sc iii}] emission as a
strong confirmation of the existence of the oxygen-bearing ``cocoon''
inferred from spectra.

The Great Eruption in the 1840's ejected about 10--15 M$_{\sun}$ of
material off the star (Smith et al.\ 2003), and $\eta$~Car may have
experienced similar mass-loss episodes before that event (Walborn et
al\ 1978).  If $\eta$~Car has a mass of $\ga$100 M$_{\sun}$
appropriate for its luminosity, then these mass ejections constitute a
significant fraction of the star's initial mass, and would have
comprised most of the star's initial outer radius.  Additionally, the
mass lost during the Great Eruption is comparable to the total mass
that was presumably lost in a normal stellar wind during the entire
time that $\eta$ Car was on the main sequence (i.e. 10$^{-6}$ to
10$^{-5}$ M$_{\odot}$ yr$^{-1}$ for $\sim$3$\times$10$^6$ yr).  Thus,
the removal of this mass in sudden outbursts may have stripped off the
remaining outer layers of the star, down to the convective core
boundary where CNO-cycle ashes could be exposed.  In that case, it is
perhaps not surprising to see a dramatic and rather sudden shift in
chemical abundances in the extensive ejecta blanket around $\eta$~Car.
The confirmation of the [O~{\sc iii}] cocoon in the images presented
here underscores our earlier speculation (Smith \& Morse 2004) about
possible connections between the change in abundances and instability
that may accompany the transition from hydrogen to helium core
burning.

The [O~{\sc iii}] veil marks the outer boundary of $\eta$ Car's
undiluted ejecta envelope.  All material inside the veil is ejecta
from $\eta$ Carinae itself, whereas outside we find ambient material
in the H~{\sc ii} region, stellar wind material from other O stars, or
both of these mixed with $\eta$~Car's ancient main-sequence wind.  In
this sense, the limb-brightened edge of the [O~{\sc iii}] veil is
probably a contact discontinuity in the wind's terminal shock --
i.e. it is the {\it astropause} of $\eta$ Car.  This is not like the
astropause for a normal star (like the heliopause) where the wind has
never reached the interstellar medium beyond.  The situation is also
unlike the standard theory for interstellar bubbles (i.e., Castor,
McCray, \& Weaver 1975), where the wind of a single massive star blows
a low-density bubble in a uniform surrounding medium, sweeping up a
dense shell.  Instead, the observed structure around $\eta$~Car is the
result of a dense slow wind expanding into a cavity created by its own
previous fast wind, and now being shaped from the outside by the fast
lower-density wind of a nearby star.  Indeed, Figure 1 gives the
impression that the less dense wind and radiation from HD~303308 are
sculpting the surrounding cavity, halting the expansion of the [O~{\sc
iii}] veil as adjacent ejecta from $\eta$~Car are swept back. This
scenario has severe asymmetries and time-dependent mass-loss
($\eta$~Car entering the LBV phase, followed by multiple eruptions)
that cannot be dealt with easily in the analytic framework of stellar
wind bubbles.  A better way to look at the problem may be from the
point of view of pressure balance between two interacting stellar
winds.

This astropause probably marks the limiting extent of a recent
enhanced heavy mass-loss phase (i.e. the LBV phase) for $\eta$~Car,
which is now being eroded from the outside, whereas $\eta$ Car's
main-sequence wind probably filled much of the surrounding H~{\sc ii}
region.  If the [O~{\sc iii}] veil were produced by $\eta$ Car's
main-sequence wind, it wouldn't have the high overdensity compared to
the adjacent wind bubble of HD~303308; indeed, on the main sequence,
$\eta$~Car was probably like the O2~If* supergiant HD~93129A (Taresch
et al.\ 1997; Simon et al.\ 1983), and the wind would have been
stronger than that of HD~303308.  Instead, for the two winds to
approximately balance with the much higher observed density on $\eta$
Car's side, it makes more sense for $\eta$~Car's wind to have filled
the interior of the [O~{\sc iii}] veil with a slower and denser wind.
The interface between the winds should occur at the ram-pressure
balance point between the two stars where $\rho_1v_1^2$=$\rho_2v_2^2$,
given by

\begin{displaymath}
\frac{R_1}{R_2}=\sqrt{\frac{\dot{M}_1 v_1}{\dot{M}_2 v_2}},
\end{displaymath}

\noindent where $R_1$ and $R_2$ are the radial separations from each
of the two stars. Using $\dot{M}_1\simeq$10$^{-3}$ M$_{\sun}$
yr$^{-1}$ and $v_1\simeq$500 km s$^{-1}$ as plausible values for
$\eta$ Car's wind,\footnote{These are roughly the {\it present day}
parameters for $\eta$~Car's wind (Hillier et al.\ 2001), which don't
necessarily apply before the Great Eruption.} and
$\dot{M}_2\simeq$10$^{-6}$ M$_{\sun}$ yr$^{-1}$ and $v_2\simeq$3100 km
s$^{-1}$ as likely values for the O4~V((f)) star HD~303308 (Repolust
et al.\ 2004), we have $R_1/R_2 \simeq$9.  This would be much closer
to HD~303308 than the observed position of the interface; the actual
observed value of $R_1/R_2\simeq1-2$ suggests that $\eta$ Car's
pre-eruption mass-loss rate was lower --- closer to 10$^{-4}$
M$_{\odot}$ yr$^{-1}$.\footnote{While this provides only
circumstantial evidence that $\eta$~Car's mass-loss rate is higher now
than it was before the Great Eruption, one can imagine why it might be
true.  The 19th-century Great Eruption removed at least 10 M$_{\odot}$
from the star (Smith et al.\ 2003), leaving $\eta$~Car with a
significantly higher L/M ratio after the outburst. The weaker gravity
as compared to before the eruption brings it $\sim$10\% \ closer to
the Eddington limit, making it easier for the radiation-driven stellar
wind to lift material off the star.}

Another reason to suppose that the astropause marks the wind from the
beginning of $\eta$ Car's LBV phase is the timescale involved.  A
coherent structure like the [O~{\sc iii}] veil would disperse when its
age exceeds the sound crossing time, which is roughly 6$\times$10$^4$
yr.  This is much shorter than the $\sim$3 Myr main sequence lifetime
for a M$_{\rm ZAMS}$=120 M$_{\odot}$ star (Chiosi \& Maeder 1986), but
comparable to the expected duration of the LBV phase (e.g., Bohannan
1997).  A mass-loss rate of perhaps 10$^{-4}$ M$_{\odot}$ during that
time (see above) would fill the volume of the [O~{\sc iii}] veil,
having a radius of roughly 30\arcsec\ or 10$^{18}$ cm, with a mass of
6~M$_{\odot}$ and an average density of about 400 cm$^{-3}$.  This is
consistent with the electron density of 470$\pm$170 cm$^{-3}$ that we
measured previously from spectra of the west edge of the veil (Smith
\& Morse 2004).  At this density, however, thermal gas pressure alone
is not enough to thwart the advance of HD~303308's wind.  This erosion
will proceeded until denser inner parts of the cocoon are reached;
this resistance would require densities inside the cocoon of

\begin{displaymath}
n_H \simeq \frac{\dot{M}_2 v_2 }{4 \pi R_2^2 k T}
\end{displaymath}

\noindent where $\dot{M}_2$, $v_2$, and $R_2$ are the values for
HD~303308's wind, as above, and $T\simeq$10$^4$ K is the temperature
of the ionized gas inside the cocoon.  For $R_1 \simeq R_2 \simeq
$10$^{18}$ cm, the density inside the cocoon should be roughly 2000
cm$^{-3}$, higher than observed at the edge of the veil (unless
$R_2$ is actually larger than the apparent separation in Figure 1 due
to a projection angle), but less than densities for some of the
[N~{\sc ii}] knots inside it (Smith \& Morse 2004).

Further observational work can potentially help clarify the nature of
the [O~{\sc iii}] veil seen in our images.  For example, high
dispersion long-slit spectra of the [O~{\sc iii}] $\lambda$5007
emission line could be used to study the kinematics of the veil and
the more extended shell.  We would expect Doppler shifts of the
[O~{\sc iii}] emission to be very different from [N~{\sc ii}] lines.
Absorption profiles of Ca~{\sc ii} and Na~{\sc i} in echelle spectra
of numerous stars surrounding $\eta$ Car might be used for the same
purpose.  If HD~303308 is inside a bubble as Figure 1 suggests, then a
detailed comparison of its absorption spectrum and those of several
nearby stars may prove very interesting.  Finally, one could determine
if the veil is a standing shock by searching for faint X-ray emission
that may be associated with it.

\acknowledgements \scriptsize

We thank Josh Walawender for assistance with the reduction of the
CTIO/MOSAIC images.  Support was provided by NASA through {\it HST}
grant GO-09775.03A to the Arizona State University and grant
HF-01166.01A to the University of Colorado from the Space Telescope
Science Institute, which is operated by the Association of
Universities for Research in Astronomy, Inc., under NASA contract
NAS~5-26555.  Additional support was provided by NSF grant AST
98-19820 and NASA grants NCC2-1052 and NAG-12279 to the University of
Colorado.

%% Table 1 - WFPC2 log
\begin{deluxetable}{lccl}
%\tabletypesize{\normalsize}
\tabletypesize{\scriptsize}
\tablecaption{{\it HST}/WFPC2 Observation Log}
\tablewidth{0pt}
\tablehead{
 \colhead{Date}	&\colhead{Filter} &\colhead{Camera} &\colhead{Exp.\ Time} }
\startdata
1993 Dec 31	&F502N	&WF3	&0.11 s			\\
1993 Dec 31	&F502N	&WF3	&2 $\times$ 4.0 s	\\
1993 Dec 31	&F502N	&WF3	&2 $\times$ 200 s	\\
2003 Aug 08	&F502N	&WF3	&2 $\times$ 1.0 s	\\
2003 Aug 08	&F502N	&WF3	&2 $\times$ 6.0 s	\\
2003 Aug 08	&F502N	&WF3	&2 $\times$ 230 s	\\
\enddata
\tablecomments{The 1993 observations are from program GO-5188 and the
2003 observations are from G0-9775.}
\end{deluxetable}

% FIGURE 1 ---------- MOSAIC O3
\begin{figure}
\epsscale{0.8}
\plotone{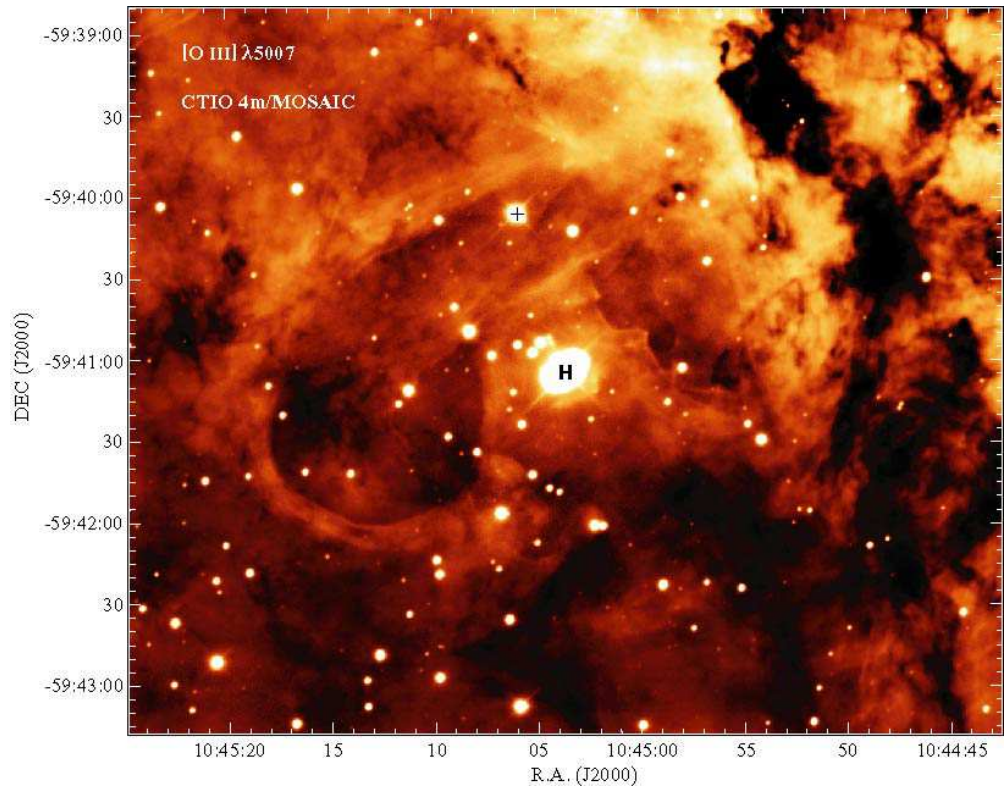}
%\plotone{nsmith.fig1color.eps}
\caption{CTIO 4m MOSAIC image of the environment around $\eta$ Car in
  the [O~{\sc iii}] $\lambda$5007 filter.  ``H'' marks the position of
  the Homunculus, and the {\bf +} marks the position of the O4 V((f))
  star HD~303308.}
\end{figure}

% Figure 2 -- WFPC2 images
\begin{figure}
\epsscale{0.95}
\plotone{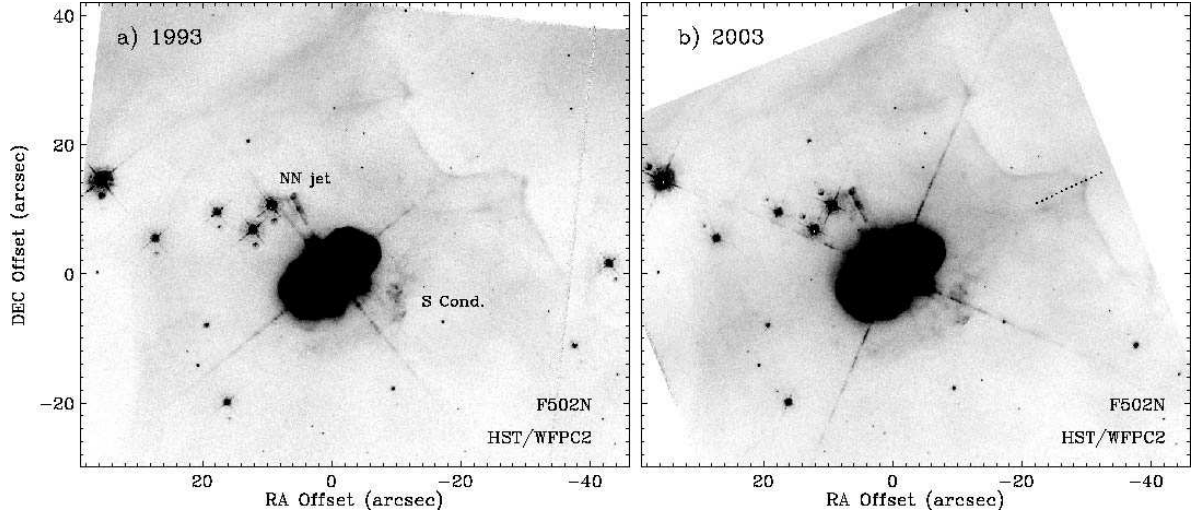}
\caption{{\it HST}/WFPC2 images of $\eta$~Car in the F502N filter
  (transmitting [O~{\sc iii}] $\lambda$5007) in (a) December 1993 and
  (b) August 2003.  The dashed line in panel (b) shows the approximate
  location of the intensity tracing through the W Arc in Figure 4.}
\end{figure}

% Figure 3 -- WFPC2 ratio image
\begin{figure}
\epsscale{0.95}
\plotone{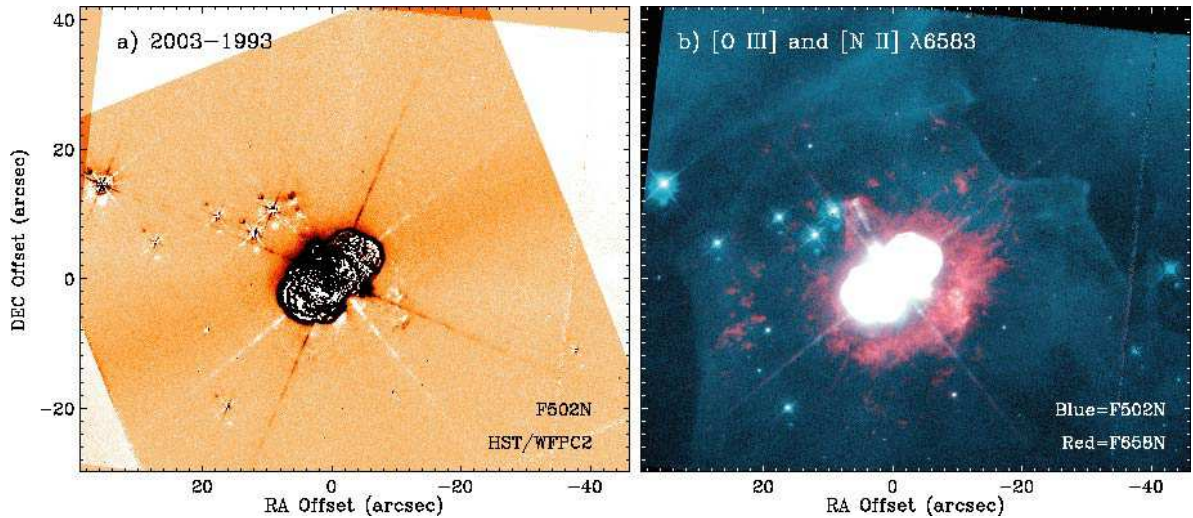}
%\plotone{nsmith.fig3color.eps}
\caption{(a) a difference image in the F502N filter, showing the
  subtraction of the two {\it HST}/WFPC2 images.  Adjacent light and
  dark residuals indicate material that has moved, while stationary
  nebular features are subtracted.  The outer [O~{\sc iii}] veil is
  not seen here because it is essentially stationary.  (b) {\it
  HST}/WFPC2 image of the outer ejecta in the F658N filter (see Morse
  et al.\ 1998), showing [N~{\sc ii}] $\lambda$6583 emission for
  comparison with the [O~{\sc iii}] emission. [N~{\sc ii}] is shown in
  red and [O~{\sc iii}] is in blue/green.}
\end{figure}

% Figure 4 -- intensity tracing of W Arc
\begin{figure}
\epsscale{0.4}
\plotone{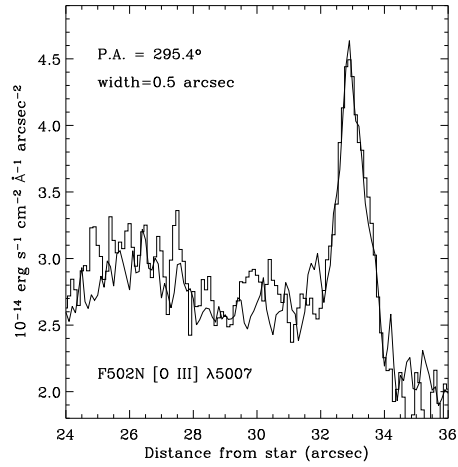}
\caption{[O~{\sc iii}] intensity tracings through the W Arc feature in
  the outer veil (see Figure 2$b$ and Smith \& Morse 2004).  The solid
  line is the [O~{\sc iii}] intensity in the F502N {\it HST}/WFPC2
  image from 1993, while the histogram is from the 2003 WFPC2 image.}
\end{figure}

\end{document}